\documentclass{modelica}

\usepackage[inline]{enumitem}

\newlist{inlinelist}{enumerate*}{1}
\setlist[inlinelist]{label=(\arabic*)}

\hypersetup{
  pdftitle  = {ModeliHub: A Web-based, Federated Analytics Platform for Modelica-centric, Model-based Systems Engineering},
  pdfauthor = {Mohamad Omar Nachawati},
  pdfsubject = {16th International Modelica Conference 2025},
  pdfkeywords = {Analytics, Federation, Model-based Systems Engineering, Unified Namespace, Virtual Twins},
  colorlinks,
  linkcolor=black,
  urlcolor=black,
  citecolor=black,
  pdfpagelayout = SinglePage,
  pdfcreator = pdflatex,
  pdfproducer = pdflatex}

\begin{document}
\thispagestyle{empty}

\title{ModeliHub: A Web-based, Federated Analytics Platform for Modelica-centric, Model-based Systems Engineering}
\author[1]{Mohamad Omar Nachawati}
\affil[1]{Independent Researcher {\small\texttt{mnachawa@gmail.com}}}

\maketitle\thispagestyle{empty}

\abstract{
This paper introduces ModeliHub, a Web-based, federated analytics platform designed specifically for model-based systems engineering with Modelica.
ModeliHub's key innovation lies in its Modelica-centric, hub-and-spoke federation architecture that provides systems engineers with a Modelica-based, unified system model of repositories containing heterogeneous engineering artifacts.
From this unified system model, ModeliHub's Virtual Twin engine provides a real-time, interactive simulation environment for deploying Modelica simulation models that represent digital twins of the virtual prototype of the system under development at a particular iteration of the iterative systems engineering life cycle.
The implementation of ModeliHub is centered around its extensible, Modelica compiler frontend developed in Isomorphic TypeScript that can run seamlessly across browser, desktop and server environments.
This architecture aims to strike a balance between rigor and agility, enabling seamless integration and analysis across various engineering domains.
}


\section{Introduction}
\label{sec:introduction}

The paradigm shift from document-based to model-based systems engineering has revolutionized how cyber-physical systems are designed and manufactured, promising a new era of efficiency and innovation.
With this shift, digital models have become the primary sources of truth, supplanting the role of textual documents in systems engineering projects.
However, despite the promise of model-based systems engineering, its benefits remain inconclusive \cite{henderson_value_2021}.

Among the challenges reported in model-based systems engineering projects is the difficulty of balancing rigor with agility \cite{tolentino_balancing_2018}.
While rigor is a crucial part of systems engineering that ensures accuracy and consistency, without which a system development project cannot honestly be called engineering, rigor can also hinder agility, as changes to a system's requirements and design can take longer complete.
This effect can have significantly more impact on early-stage systems engineering projects where the requirements and design of a system remain in flux.

Compounding the difficulty of balancing rigor with agility is the lack of a single source of truth in heterogeneous model-based systems engineering projects.
These projects often involve the use of different tools and languages to model various aspects of the system under development, such as requirements, architecture, computer-aided design and engineering, dynamic simulation, and analysis.
Under these conditions, traceability links need to be maintained not only across different modeling elements within a single tool or language but also across all tools and languages used throughout the project.
Digital thread platforms, like Syndeia \cite{bajaj_architecture_2016}, aim to address this difficulty by providing adapters for tools and languages commonly used in systems engineering projects.
These adapters facilitate the creation and maintenance of traceability links across different modeling elements across multiple tools and languages.

Researchers have also investigated the use of the Web Ontology Language \cite{horrocks_shiq_2003} to support integration and interoperability in model-based systems engineering projects.
For instance, the Industrial Ontologies Foundry (IOF) Ontology \cite{kulvatunyou_industrial_2018} aims to facilitate cross-system data integration in digital manufacturing, using \textcite{smith2013classifying}'s Basic Formal Ontology (BFO) as its top-level ontology.
SysML2's, the latest version of the Systems Modeling Language, alignment with OWL2 further demonstrates this trend.

However, while OWL2's open world assumption facilitates agility, it has limited potential for model verification purposes, which negatively impacts rigor \cite{elaasar_opencaesar_2023}.
To address this limitation, the Ontological Modeling Language (OML)\footnote{URL: http://www.opencaesar.io/oml/} serves as a layer of abstraction over OWL2 that seamlessly enables closed-world reasoning through the use of its \textit{bundle closure algorithm}\footnote{URL: https://github.com/opencaesar/owl-tools/blob/master/owl-close-world/README.md}.
This algorithm is used during the compilation of OML into OWL2 to add appropriate disjointness axioms needed to mimick the behavior of closed-world reasoning with OWL2 DL reasoners.

While previous experience during the development of GitWorks \cite{nachawati_towards_2022} confirmed the utility of translating Modelica and other artifacts into RDF triples and loading those triples into a triple store for ad-hoc querying using SPARQL\footnote{URL: https://www.w3.org/TR/sparql11-overview/}, it also highlighted the limitations of using OML for semantic analysis and verification of Modelica models.
Although certain entailments provided by a DL reasoner made SPARQL querying easier, such as the transitive closure of superclasses, complex but widely used constructs of the Modelica language, such as modifiers and redeclarations, could not be directly expressed in OML.
Furthermore, even though the OML representation of the Modelica model after flattening could be used, the Modelica compiler frontend had already effectively computed these entailments at that point.
Consequently, going directly from Modelica to RDF, bypassing OML compilation to OWL2 and the reasoning process, seemed more efficient.

This paper focuses on addressing those concerns.
Specifically, the contributions are twofold:
First, ModeliHub is introduced as a web-based, federated analytics platform designed specifically for model-based systems engineering with Modelica.
ModeliHub's key innovation lies in its Modelica-centric, hub-and-spoke federation architecture, which provides systems engineers the ability to construct analytical views and reports using a Modelica-based, unified system model of repositories containing heterogeneous engineering artifacts.
This aims to strike a balance between rigor and agility, enabling seamless integration and analysis across various engineering domains.
From this unified system model, ModeliHub's Virtual Twin engine provides a real-time, interactive simulation environment for deploying Modelica simulation models that represent digital twins of the virtual prototype of the system under development at a particular iteration of the iterative systems engineering life cycle.

Second, a prototype of ModeliHub is implemented, centered around the development of an extensible, Modelica compiler frontend written in Isomorphic TypeScript. This frontend can run seamlessly in both browser and Node.js environments.
The Modelica compiler frontend is used to populate an Oxigraph triple store, which also runs in both browser and Node.js environments, with an RDF representation of Modelica models inspired by ModelicaOML\footnote{URL: https://github.com/OpenModelica/ModelicaOML} along with entailments that would otherwise be left for reasoners to infer.
By representing non-Modelica artifacts as Modelica through the use of adapters in the extensible compiler frontend, verification of non-Modelica artifacts can be done using the frontend instead of via OWL2 and perhaps SHACL.
While Modelica \cite{fritzson_modelicaunified_1998} is primarily known for modeling and simulating cyber-physical systems, non-simulation applications like parsing and semantically analyzing Modelica source code have been demonstrated, as seen in the work of \textcite{tiller_parsing_2003} and \textcite{johansson_modelicadb-tool_2005}.

The rest of this paper is organized as follows:
Section~\ref{sec:platform-overview} provides an overview of the ModeliHub platform for Modelica-centric, model-based systems engineering, describing its architecture and showcasing some use cases of the platform, adapted from \textcite{friedenthal_architecting_2017}.
Next, Section~\ref{sec:implementation} describes the implementation of ModeliHub, focusing on the Isomorphic TypeScript-based Modelica compiler frontend.
Finally, Section~\ref{sec:conclusions-and-future-work} concludes the paper with a brief discussion of potential directions for future research and development.

\section{Platform Overview}
\label{sec:platform-overview}

This section provides an overview of the ModeliHub platform for model-based systems engineering.
It presents a high-level architecture of ModeliHub, illustrated in Figure~\ref{fig:platform-architecture}, highlighting the major components of the platform.
Then it showcases some use cases of the platform, adapted from \textcite{friedenthal_architecting_2017}.

\subsection{Architecture}

\begin{figure}[h!]
	\centering
	\includegraphics[width=\columnwidth]{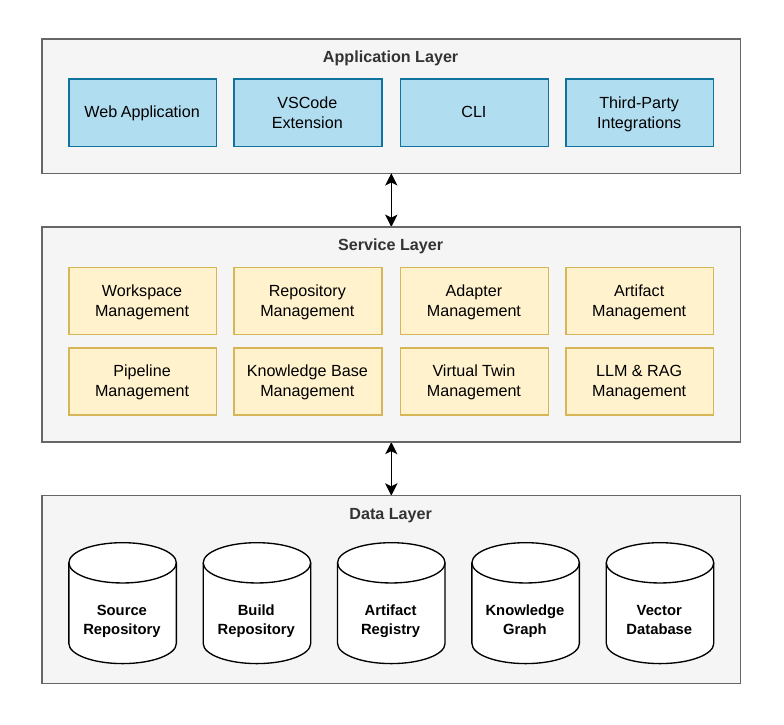}
	\caption{ModeliHub platform architecture}
	\label{fig:platform-architecture}
\end{figure}

The high-level architecture of ModeliHub, as illustrated in Figure~\ref{fig:platform-architecture}, is divided into three distinct layers: the application layer, the service layer, and the data layer.
ModeliHub provides a VSCode Web extension for authoring Modelica-based analytical views.
This extension can be used in both browser-based and desktop versions of VSCode.
ModeliHub Workspaces provides a Git-based repository for various engineering artifacts like geometric models, simulation models, reports, visualizations, and documentation.
The workspace also provides a CI environment for executing pipelines to transform artifact sources into builds.
In addition, the workspace also provides a ad-hoc queryable, graph database that is automatically populated with Modelica entities extracted from the repository.

ModeliHub features an artifact registry where engineering artifacts can be 
published and reused across different ModeliHub workspaces, adhering to the 
CommonJS package specification.
This registry encompasses a wide range of information work products generated 
throughout the product lifecycle, including CAD models, CAE studies, and their 
analyses.  A single ModeliHub workspace can publish multiple artifacts to this 
registry and simultaneously depend on other artifacts from it.
While Impact \cite{tiller_impact-modelica_2014}, a package manager for Modelica, 
already exists, ModeliHub chose to align with the JavaScript ecosystem for its 
artifact management system. 

\begin{figure}[h!]
	\centering
	\includegraphics[width=.8\columnwidth]{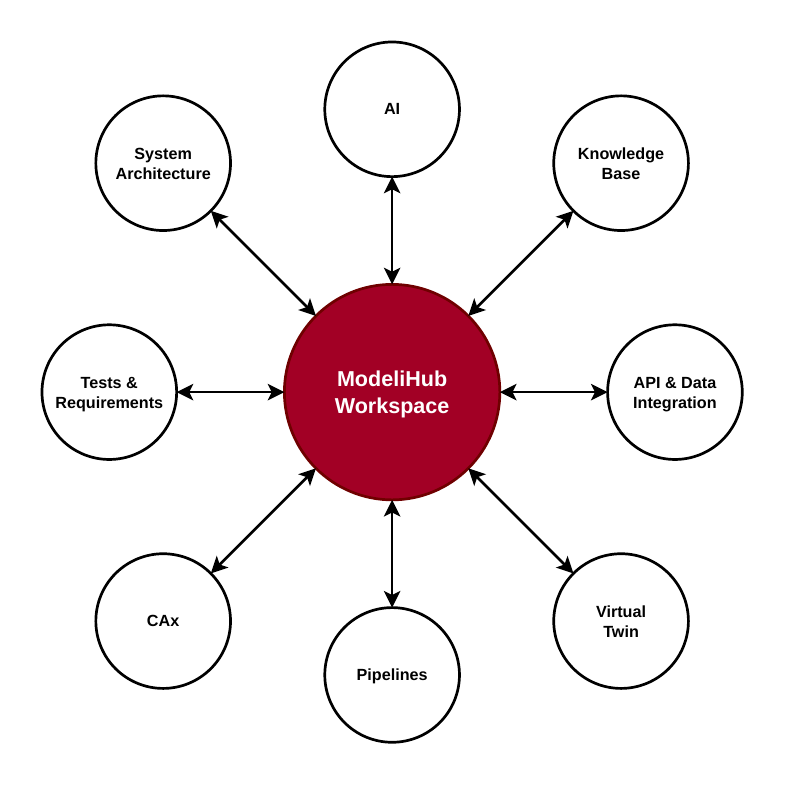}
	\caption{ModeliHub workspaces}
	\label{fig:workspaces}
\end{figure}

Conceptually, a \textit{virtual twin} is the digital twin of a virtual prototype of a system at a particular iteration of an agile systems engineering life cycle.
The \textit{unified namespace} (UNS), a term coined by Walker Reynolds\footnote{See https://inductiveautomation.com/resources/article/uns-unified-namespace}, serves as the real-time, single source of truth for the digital enterprise.
The virtual twin is represented by running a Modelica simulation in realtime and providing an MQTT interface for interacting with simulation variables.
The virtual twin management API provides support for deploying, undeploying, starting, stopping and interacting with Modelica-based simulations representing the virtual twin.

\subsection{Use Cases}

Let's consider the initial system mass budget for the spacecraft case study presented in \textcite{friedenthal_architecting_2017}.
Early in system development, these parameters might be stored in an Excel spreadsheet, potentially saved as a CSV file within the repository (as illustrated in \autoref{lst:mass-budget-csv}) or even as a JSON file (see \autoref{lst:mass-budget-json}).

\begin{lstlisting}[caption=CSV Initial system mass budget,label=lst:mass-budget-csv]
subsystem,initial-mass,margin,budget-mass
Payload,18.2,7.8,26.0
Structure,16.1,6.9,23.0
...
\end{lstlisting}

As the system development evolves, these parameters might be used in multiple engineering artifacts, e.g. the CAD model and simulation model, and could ultimately undergo many iterations of refinement.
This evolution means the initial Excel spreadsheet is no longer the single source of truth.
To maintain accuracy and rigor, these parameters must be tracked and synchronized across different tools and languages, which can significantly impede agility.

\begin{lstlisting}[caption=JSON Initial system mass budget,label=lst:mass-budget-json]
[
  {
    "subsystem": "Payload",
    "initial-mass": 18.2,
    "margin": 7.8,
    "budget-mass": 26.0
  },
  {
    "subsystem": "Structure",
    "initial-mass": 16.1,
    "margin": 6.9,
    "budget-mass": 23.0
  },
  ...
]
\end{lstlisting}

To support this scenario, the ModeliHub Modelica-centric core is extended with adapters that can read files in various formats and automatically convert them into virtual Modelica fragments.
These fragments are then fed into the Modelica language server and compiler, allowing seamless interaction with other Modelica components as if the parameters were originally encoded in a concrete Modelica file.
This adaptive approach diverges from traditional data retrieval APIs, enabling a more dynamic and integrated workflow within ModeliHub's virtual twin-based systems engineering environment.
\autoref{lst:virtual-modelica} provides an example of a virtual Modelica fragment generated automatically from the CSV file in \autoref{lst:mass-budget-csv} and the JSON file in \autoref{lst:mass-budget-json}.

The package name and namespace are automatically inferred from the file's location and name.
The package declares a type as well as a constant instance by inferring the structure from the JSON or CSV file.
Subsequently, this file can be imported by and used in other Modelica components, treating these parameters as if they were encoded in a concrete Modelica file.
This approach is markedly different from the mechanism proposed by \textcite{tiller_implementation_2005} that involved a data retrieval API called from within the model.

\begin{lstlisting}[caption=Virtual Modelica representation of CSV file,label=lst:virtual-modelica]
within acme.engine;
package MassBudget
  record MassBudget
    String subsystem;
    Real   initialMass;
    Real   margin;
    Real   budgetMass;
  end MassBudget;
  constant MassBudget root[:] = {
    Test(
      subsystem   = "Payload",
      initialMass = 18.2,
      margin      = 7.8,
      budgetMass  = 26.0
    ),
    Test(
      subsystem   = "Payload",
      initialMass = 18.2,
      margin      = 7.8,
      budgetMass  = 26.0
    ),
    ...
  };
end MassBudget;
\end{lstlisting}

ModeliHub analytical views are implemented using JavaScript. These Modelica 
analytical views leverage access to the Modelica class tree, instance tree, and 
an indexed, RDF representation of the instance tree.
ModeliHub provide built-in rendering formats for these analytical views, including:
\begin{itemize}
  \item A diagram view powered by the X6 JavaScript Diagramming Library from the AntV group\footnote{URL: https://x6.antv.dev/}
  \item A graph view based on Cytoscape.js\footnote{URL: https://js.cytoscape.org/}
  \item A table view utilizing AG Grid\footnote{URL: https://www.ag-grid.com/}
  \item A geometry view leveraging Three.js\footnote{URL: https://threejs.org/}
\end{itemize}

\begin{figure}[h!]
	\centering
	\includegraphics[width=.8\columnwidth]{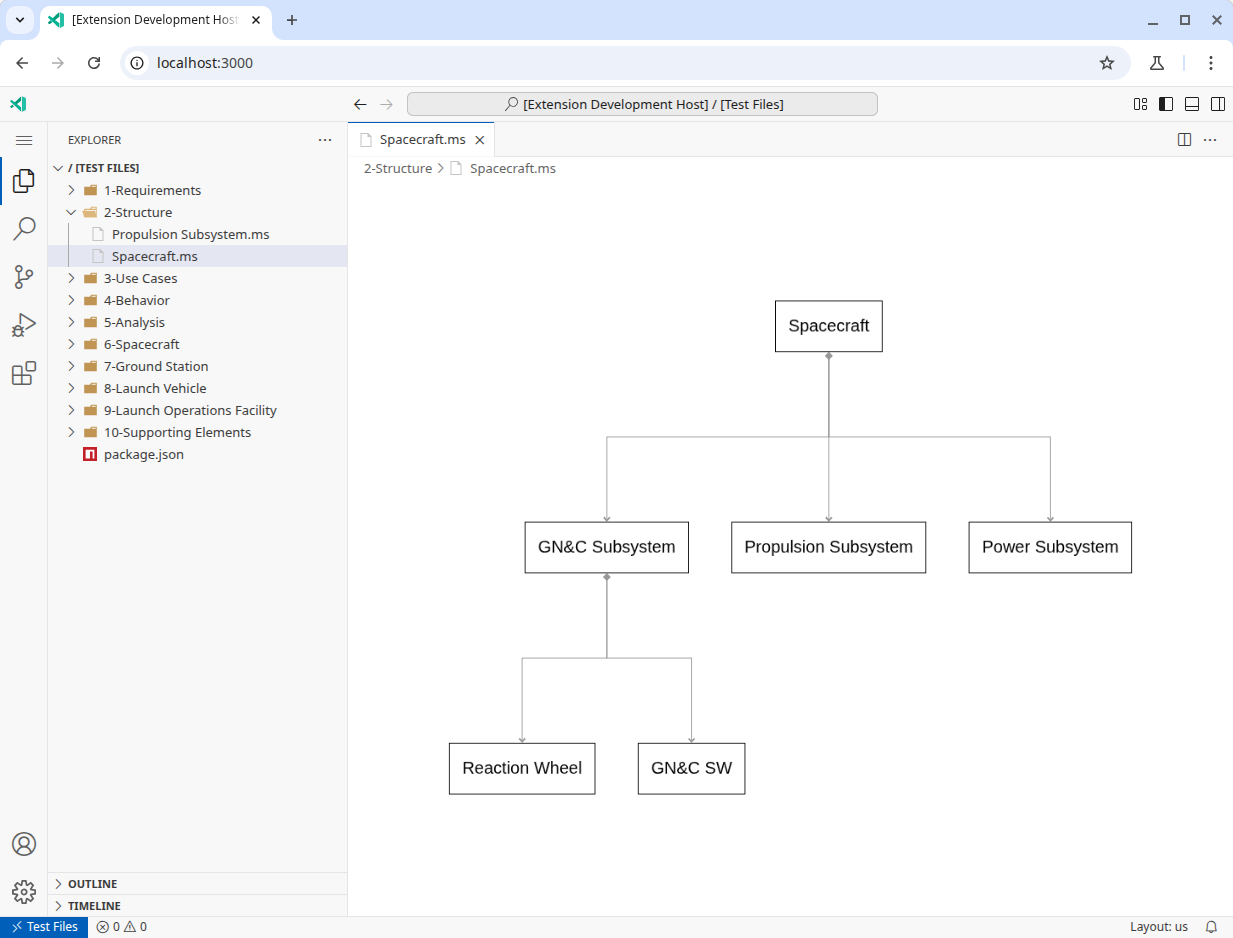}
	\caption{Composition view}
	\label{fig:composition-view}
\end{figure}

The Modelica language provides constructs for modeling system architectures.
Modelica's diagrammatic representation closely mirrors the interconnection view found in SysML, visually highlighting the internal structure and connections of a system.
ModeliHub extends these capabilities to offer a visual representation of system composition, inheritance hierarchies, along with interconnections between components.
This enables engineers to effectively grasp and manage complex systems by providing intuitive insights into their structure and relationships.

\begin{figure}[h!]
	\centering
	\includegraphics[width=.8\columnwidth]{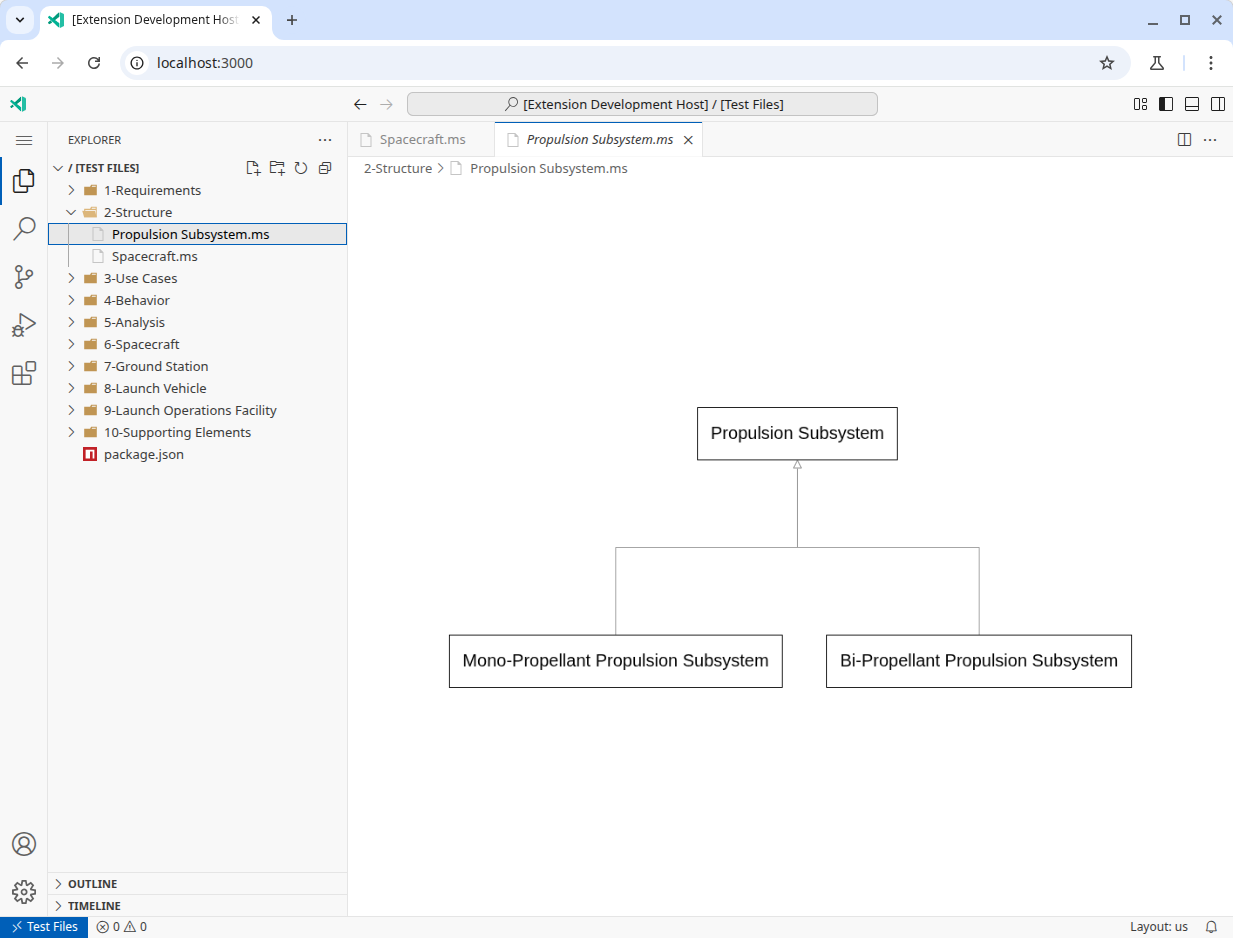}
	\caption{Specialization view}
	\label{fig:specialization-view}
\end{figure}

Capturing and managing requirements are paramount in systems engineering.
ModeliHub provides tabel-based requirements views that offer a comprehensive overview of system requirements, enabling easy association with specific parts of the unified system model.

Synchronizing geometry with system architecture and requirements is a key objective of digital thread platforms. ModeliHub facilitates this by integrating 3D geometric views. Data from various artifacts within the system model can be overlaid onto these geometries, providing a cross-cutting perspective and streamlining navigation to related information.

\begin{figure}[h!]
	\centering
	\includegraphics[width=.8\columnwidth]{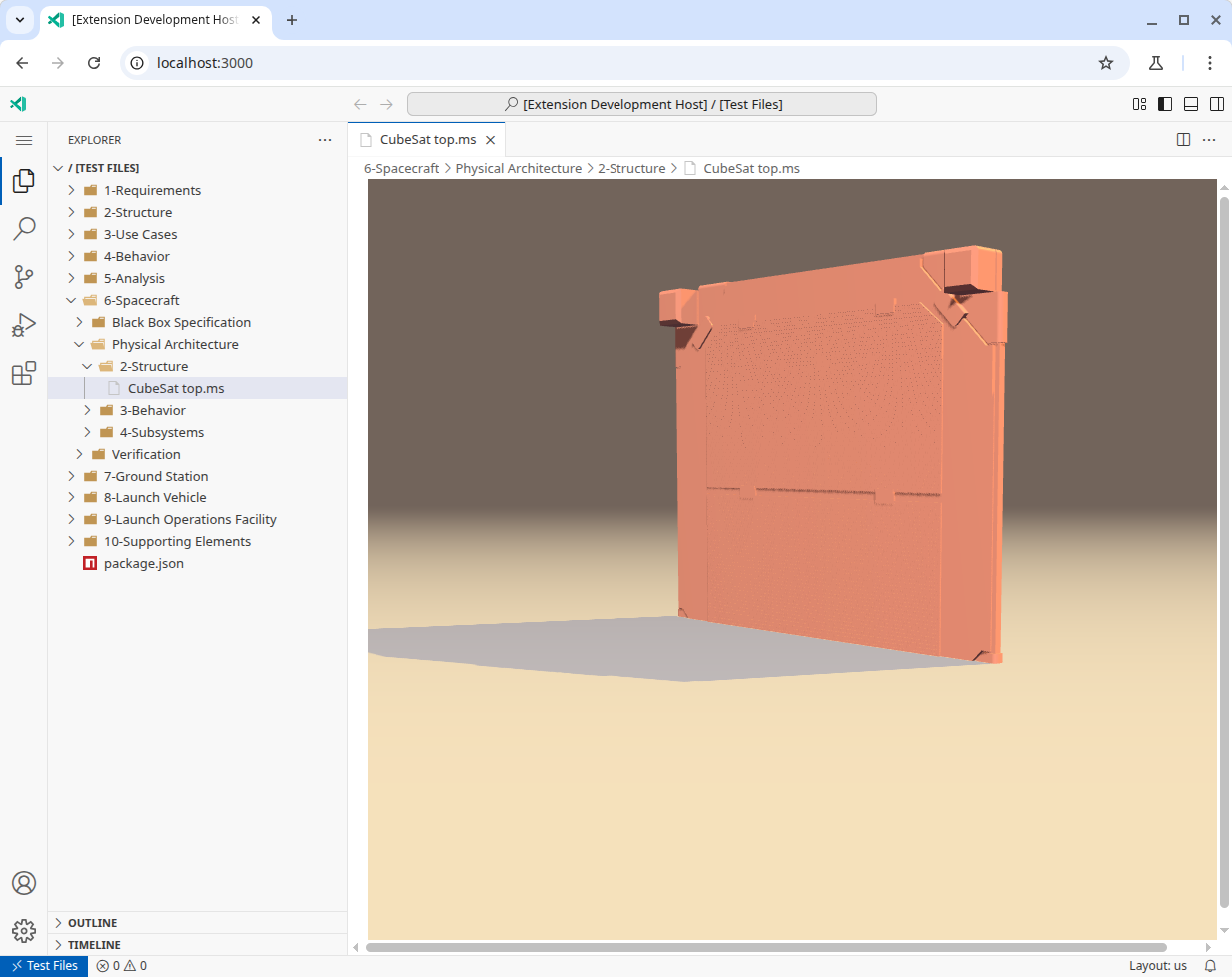}
	\caption{Geometry view}
	\label{fig:geometry-view}
\end{figure}

ModeliHub's knowledge base pages facilitate the development of system documentation that's always in sync with the unified system model.
It does this by integrating both the Modelica compiler frontend and Markdown files stored within the repository.
This knowledge base enables dynamic expressions embedded within text blocks, ensuring that documentation is always up-to-date and accurately reflects any changes made to the system model.

\section{Implementation}
\label{sec:implementation}

This section details the implementation of ModeliHub, focusing on its extensible Modelica compiler frontend developed using Isomorphic TypeScript.

\subsection{Modelica Compiler Frontend}

ModeliHub's core functionality lies in its ability to process Modelica source code in both browser and server environments.
This is achieved through the development of a Tree-Sitter grammar for Modelica 3.6, allowing for parsing of Modelica text into a syntax tree.
Unlike for batch compilation processes, the language tooling for IDEs and interactive tools need to be able to synchronize changes between the textual representation and the parse tree.
While Tree-sitter can incrementally generate and maintain a concrete syntax tree representation of the source text, the concrete syntax tree generated by Tree-Sitter is not directly suitable for instantiation algorithm or other functions operating on the parse tree.
The concrete syntax tree is therefore transformed into an abstract syntax tree (AST) more amenable to analysis and manipulation.

The ModeliHub compiler frontend is designed to be extended with adapters for reading and writing non-Modelica files as if they were written in Modelica.
For example, the CSV adapter makes CSV files available as a Modelica array of class instances.
The field names of the class are derived from the first line of the CSV files, and the field types are inferred by analyzing the data in the proceeding lines.
Similarly, the JSON adapter makes JSON files available as Modelica instances or arrays.
While the frontend currently supports a subset of the full specification, development is ongoing, with the aim of supporting the full language specifciation. Currently the frontend supports modifiers, redeclarations, annotations, function calls, and a significant porition of array processing.

\subsection{Knowledge Graph}

The Modelica compiler frontend is used to populate an Oxigraph triple store, which also runs in both browser and Node.js environments, with an RDF representation of Modelica models inspired by ModelicaOML\footnote{URL: https://github.com/OpenModelica/ModelicaOML} along with entailments that would otherwise be left for reasoners to infer.
By representing non-Modelica artifacts as Modelica through the use of adapters in the extensible compiler frontend, verification of non-Modelica artifacts can be done using the frontend instead of via OWL2 and perhaps SHACL.
Oxigraph's SPARQL query engine allows for powerful and flexible querying of the knowledge graph, enabling users to extract specific information or insights from the model in an analytical view.

\subsection{Virtual Twin Engine}

\begin{figure}[h!]
	\centering
	\includegraphics[width=.8\columnwidth]{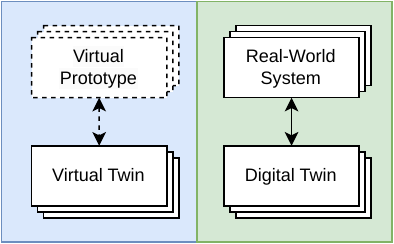}
	\caption{Virtual twins}
	\label{fig:virtual-twin}
\end{figure}

The ModeliHub Virtual Twin engine provides an interactive, real-time runtime environment for deploying Modelica simulation models as a Unified Namespace\footnote{See https://inductiveautomation.com/resources/article/uns-unified-namespace} over MQTT.
MQTT is a popular, lightweight IOT messaging protocol.
As shown in \autoref{fig:virtual-twin}, a virtual twin is the digital twin of a virtual prototype of a system at a particular iteration of an iterative systems engineering life cycle.

ModeliHub's Virtual Twin engine enables real-time simulation and interaction with Modelica models.
Ultimately, the vision is to develop the Isomorphic TypeScript Modelica compiler frontend into a full Modelica compiler that supports seamless interactive simulation of Modelica models in both the browser and the desktop.
In the meantime, ModeliHub leverages OpenModelica's \cite{fritzson_openmodelica_2005} capabilities for embedded OPC UA servers and real-time simulation. 

OpenModelica provides simulation flags to enable an embedded OPC UA server and to run simulation models in realtime. 
The Virtual Twin engine uses Docker to encapsulate a Virtual Twin instance that uses OpenModelica with the flags \texttt{-rt=1.0} and \texttt{-embeddedServer=opc-ua} to run the simulation in realtime.
Furthermore, the Virtual Twin engine provides an OPC UA-MQTT gateway to translate OPC UA API from OpenModelica into MQTT and MQTT over WebSocket for communcation with ModeliHub clients.
The clients use MQTT.js to communicate with the Virtual Twin engine.

\section{Conclusions and Future Work}
\label{sec:conclusions-and-future-work}

ModeliHub has been introduced as a Web-based, federated analytics platform for model-based systems engineering that balances between rigor and agility, enabling seamless, Modelica-centric integration and analysis across various engineering domains.
ModeliHub's key innovation lies in its Modelica-centric, hub-and-spoke federation architecture that provides systems engineers with a Modelica-based, unified system model of repositories containing heterogeneous engineering artifacts.
ModeliHub's implementation leverages an extensible, Modelica compiler frontend developed in Isomorphic TypeScript that can enables seamless integration and analysis across browser, desktop and server environments.

The development of ModeliHub is an ongoing effort.
Future directions for research and development of ModeliHub include:
\begin{inlinelist}
\item development of adapters to support the Modelica-based federation of additional engineering artifact formats and protocols,
\item development and maintenance of ModeliHub's Isomorphic TypeScript-based Modelica compiler frontend and its extension with backend algorithms (e.g. matching, sorting, index reduction and tearing algorithms) and solvers to enable the dynamic analysis of Modelica-based unified system models in the browser, and
\item research on surrogate-based methods to accelerate the dynamic analysis of large and complex models, perhaps along the lines of \textcite{nachawati_mixed-integer_2021}.
\end{inlinelist}

\printbibliography

\end{document}